\documentclass[showpacs,nofootinbib,eqsecnum]{revtex4}
\usepackage{graphicx}
\usepackage{latexsym}

\def\be{\begin{eqnarray}}
\def\ee{\end{eqnarray}}
\usepackage{amsmath}
\usepackage{amsfonts}
\usepackage{amssymb}
\usepackage{epsfig}
\usepackage{color}
\usepackage{mathrsfs}
\usepackage{indentfirst}

\definecolor{dyellow}{rgb}{1.,0.8,.0}
\definecolor{myblue}{rgb}{.1,.1,.7}
\definecolor{dcyan}{rgb}{.0,.6,.6}
\definecolor{dmagenta}{rgb}{0.6,0.0,0.6}
\definecolor{brown}{rgb}{0.6,0.2,0.}
\definecolor{darkblue}{rgb}{.0,.0,0.5}
\definecolor{darkred}{rgb}{0.75,0.0,0.0}
\definecolor{orange}{rgb}{1.,.6,.0}
\definecolor{dorange}{rgb}{0.8,.4,.0}
\definecolor{darkgreen}{rgb}{0.0,0.6,0.0}
\definecolor{purple}{rgb}{.4,.0,.4}

\def\al{\alpha}

\def\dl{\delta}

\def\la{\lambda}

\def\Om{\Omega}

\def\bc{\begin{center}}
\def\ec{\end{center}}
\def\be{\begin{eqnarray}}
\def\ee{\end{eqnarray}}

\newcommand{\omits}[1]{}

\def\d#1#2{\displaystyle\frac{\displaystyle #1}{\displaystyle #2}}
\def\r{\partial}

\newcommand\btd{\raise 2pt
\hbox{$\hat\bigtriangledown$}\hskip 1.5pt}
\newcommand\bt{\raise 2pt
\hbox{$\bigtriangledown$}\hskip 1.5pt}

\def\PRD{{\it Phys. Rev.}~{\bf D}}
\def\PRL{{\it Phys. Rev. Lett }}

\def\PLB{{\it Phys. Lett.}~{\bf B}}

\def\RMP{{\it Rev. Mod. Phys. }}

\def\CTP{{\it Commun. Theor. Phys. }}
\begin{document}

\bigskip

\title{A new single-dynamical-scalar-field model of dark energy}

\author{Chao-Guang Huang}
\affiliation{Institute of High Energy Physics, Chinese Academy
of Sciences, P.O. Box 918-4, Beijing 100049, China}
\author{Han-Ying Guo}
\affiliation{Institute of Theoretical Physics, Chinese Academy
of Sciences, P.O. Box 2735, Beijing 100080, China}

\begin{abstract}
A new single-dynamical-scalar-field model of dark energy is proposed, in which 
either higher derivative terms nor structures of extra dimension are 
needed.  With the help of a fixed background vector field, the 
parameter for the effective equation of state of dark energy may cross 
$w=-1$ in the evolution of the universe.  After suitable choice of the
potential, the crossing $w=-1$ and transition from decelerating to 
accelerating occur at $z\approx 0.2$ and $z\approx 1.7$, respectively. 
\end{abstract}


\maketitle

\tableofcontents

\newpage

\section{Introduction}

Recent observations\cite{Riess98,Perl99,WMAP} suggest that the 
universe consists of dark energy ($73\%$), dark matter ($23\%$) 
and baryon matter ($4\%$).  To understand the nature of the dark energy 
is one of essential tasks in cosmology.  The most simple form of 
the dark energy is the cosmological constant, which fits well
to the observation data \cite{seljak} and which has the effective
equation of state $p=-\rho$, {\it i.e.} $w\equiv -1$.  If the dark energy is
totally contributed by the cosmological constant, the universe
is asymptotically de Sitter one and we should study the 
de Sitter spacetime and asymptotically de Sitter spacetimes in
all aspects, including in the framework of general relativity \cite{dSgr},
in the framework of de Sitter invariant special relativity \cite{dSsr1},
and locally de Sitter invariant gauge theory of gravity \cite{dSlocal,
dSlocal2}, etc.

However, there is an evidence  \cite{ex} to show that the dark energy 
might evolve from $w>-1$ in the past to $w<-1$ today and cross $w=-1$ 
in the intermediate redshift.  Obviously, the simple dynamical dark 
energy models considered vastly in the literature as well as the 
cosmological constant cannot explain it.  For example, $w$ in
the quintessence models \cite{quintessence} 
is always evolving  in the range of $-1 \leq w \leq 1$. In the
phantom models \cite{phantom}, regardless its wrong-sign kinetic 
energy term, $w$ always varies below $-1$.  The parameter $w$ of 
the effective equation of state for the 
general k-essence models \cite{vikman} also fails to cross $w=-1$.
Recently, many dark energy models with the property of crossing $w=-1$ have 
been proposed \cite{quintom,lfz,extra-d,ZY-AP,DeDeo}.  In the dynamical-scalar-field models of dark energy \cite{quintom,lfz},
either the malformed phantom fields or higher derivative term are needed.  
In \cite{extra-d}, the structures in extra dimension are introduced to realize
the parameter $w$ crossing $-1$.  In \cite{ZY-AP, DeDeo}, the dark energy is supposed to be contributed from a vector field \cite{ZY-AP} or a gas of
fermion particles interacting with a condensate represented by a vector field $b_\mu$ \cite{DeDeo}.  The vector-field and fermion-field models of the dark energy do not need phantom fields, higher derivative terms and more complex structure in extra dimension and so on.

In the present paper, 
we shall propose a new single-dynamical-scalar-field model, as a toy, of dark
energy, which may explain that the parameter $w$ crosses $-1$ at $z\approx 0.2$ and the the universe experience a phase transition form decelerating to
accelerating at $z\approx 1.7$.   
The basic idea is that the dark energy is not direct effect of 
a fundamental field, while it is only effectively described by a 
dynamical scalar field.   Taking it into account, we may consider the 
following model of a single dynamical scalar field  
coupled to an {\it a priori} non-dynamical, background covariant vector field,
%
\be\label{action}
S_\phi^{\rm eff}=\int d^4x \sqrt{-g}\left ({1\over 2}g^{\mu \nu}\phi_{,\mu}\phi_{\nu}-g^{\mu \nu}\phi_{,\mu} A_\nu-V(\phi)\right)
\ee
%
where the vector $A_\mu$ describes some unknown effects of the universe and 
is supposed to have 
constant zeroth-component in a comoving system. 
In such a model, no phantom is needed to realize $w<-1$ and 
$w$ crossing $-1$.

In the following, we shall introduce the model first. And then we shall 
make some numerical analysis with an
ordinary 
potential and a potential inspired from supergravity \cite{BM}.  Finally, we shall give some concluding 
remarks.

\section{A new single-dynamical-scalar field model}

Suppose that the universe is governed by the action
\be
S=S_{\rm EH}+S_{\rm fluid}+S_\phi^{\rm eff},
\ee
where $S_{\rm EH}$ is the Einstein-Hilbert action of gravitation, 
$S_{\rm fluid}$ is the action of perfect fluid, describing the baryonic matter
and cold dark matter, and $S_\phi^{\rm eff}$ is the effective action of 
dark energy, which has been given in Eq.(\ref{action}).  Since $A_\mu$ is introduced as a part of the effective model of dark energy,
we suppose that it only couples to the scalar field but does not couple to 
the matter.  As the standard treatment in cosmology, we suppose
that the universe is homogeneous and isotropic on large scale, and further suppose that the universe is flat for simplicity, so that the geometry of
the universe  is described by the Robertson-Walker metric
\be \label{RWmetric}
ds^2=dt^2-a^2(t)\left ({dr^2} + r^2 d\Om_2^2 \right ),
\ee
where $d\Om_2^2$ is the line-element on a unit sphere.  Under the assumption of homogeneity and isotropy, all quantities, including $\rho$, $\phi$, etc., depend on $t$ only.  In particular, the stress-energy tensor for the scalar field, defined by 
\be  \label{stress} 
T_{\mu\nu}= \d 2 {\sqrt{-g}}\d {\dl S_\phi^{\rm eff}}{\dl g^{\mu\nu}},
\ee
reads
\be \label{stress-cp}
T_{\mu\nu}= \dot \phi ^2(t) \dl^0_\mu \dl^0_\nu 
-2\dot \phi (t)  A_0 \dl^0_\mu \dl^0_\nu - \d 1 2 g_{\mu \nu}\left (
\dot \phi ^2(t) -2\dot \phi (t)A_0 -2V[\phi(t)]\right).
\ee
Comparing the stress-energy tensor (\ref{stress-cp}) with the one for 
perfect fluid, one may define the effective energy density and pressure 
of scalar field by 
\be
\rho_\phi^{}(t) &=& T_{00}=\d 1 2 \dot \phi ^2(t) -\dot \phi (t)  A_0  +  
 V[\phi(t)], \\
p_\phi^{}(t) &=& -g^{11}T_{11}= \d 1 2 
\dot \phi ^2(t) -\dot \phi (t) A_0 -V[\phi(t)].
\ee
As the effective description of dark energy, $\rho_\phi^{} =\frac 1 2\dot \phi (\dot \phi -2 A_0)+V$ should always be positive.
The parameter $w$ in the effective equation of state $p_\phi^{}=w\rho_\phi^{}$ is then
\be \label{w}
w= \d {\dot \phi ^2 -2\dot \phi A_0 -2V(\phi)}{\dot \phi ^2 -2\dot \phi  A_0  +2  
 V(\phi) }. 
\ee
It is obvious that for $A_0>0$,  
\be
w\begin{cases}  \geq -1, & \dot \phi \geq  2A_0 \quad {\rm or}\quad \dot \phi \leq 0, \\
                < -1, & 2A_0 > \dot \phi >0  \quad {\rm and} \quad \dot\phi^2-2A_0\dot \phi + 2V(\phi) >0, \\
\mbox{irrelevant}, & 2A_0 > \dot \phi >0  \quad {\rm and} \quad \dot\phi^2-2A_0\dot \phi +2V(\phi) \leq 0,
\end{cases}
\ee
and for $A_0<0$, 
\be 
w\begin{cases} \geq -1, & \dot \phi \geq 0 \quad {\rm or}\quad \dot \phi \leq 2A_0, \\
< -1, & 0 > \dot \phi > 2A_0  \quad {\rm and} \quad \dot\phi^2-2A_0\dot \phi +2V(\phi) >0, \\
{\rm irrelevant}, & 0 > \dot \phi > 2A_0  \quad {\rm and} \quad \dot\phi^2-2A_0\dot \phi +2V(\phi) \leq 0.
\end{cases}
\ee
Namely, the parameter $w$ may cross $-1$ in the evolution of the universe.

As mentioned in the previous section, we suppose that $A_\mu$
has a constant norm and constant zeroth-component in Robertson-Walker
metric. 
Under the assumption, the equation of motion for $\phi(t)$ 
takes the form
\be \label{EoMphi}
\ddot \phi +3H\dot \phi +\d {\r V}{\r \phi} = 3HA_0,
\ee
where $H=\dot a/a$ is the Hubble parameter.
Comparing with the standard form of the equation of motion for a scalar field, there exists an additional current due to the existence of $A_0$.
Now, the Friedmann equation reads
\be \label{Friedmann1}
H^2 = \d {8\pi G} 3\left (\d 1 2 \dot \phi ^2 -\dot \phi  A_0  +  
 V(\phi) + \rho_{\rm fluid}^{}  \right ), 
\ee
where the last term on the right-hand-side is the energy density of 
matter as a perfect fluid.  In the present paper, we are only interested 
in the late stage of the evolution of the universe, in which both the 
baryon matter and cold dark matter can be treated as pressureless perfect 
fluid.  Thus, we may set
\be
\rho =\rho_0^{}a_0^3/a^3,
\ee 
where $\rho_0^{}$ is a constant, denoting the present value of the 
energy density of all matter.  The subscript 0, except in $A_0$ and the 
following $b_0$, represents to take the value today.

\section{Numerical analysis}

In order to solve Eqs.(\ref{EoMphi},\ref{Friedmann1}) numerically, we recast them as the first-order differential equations with respect to the redshift 
$z=(a_0/a)-1$,
\be
\d {d \varphi}{dz} &=& - \d \chi {h (1+z)} \\
\d {d\chi}{dz} &=&3\d {\chi-b_0}{1+z}+ \d {v'(\varphi)}{h(1+z)} \\
h^2&=&\d 1 2 \chi^2 -b_0\chi +v(\varphi) +\Om_{M0}^{}(1+z)^3 ,
\ee
where 
\be
\varphi = \sqrt{\d {8\pi G} 3}\phi , \qquad  \chi =  \sqrt{\d {8\pi G} 3} \d {\dot \phi} {H_0^{}}, \qquad  b_0 = \sqrt{\d {8\pi G} 3} \d {A_0^{}} {H_0^{}}, \ee
\be
h=\d H {H_0}, \qquad v(\varphi)=\d {8\pi G V(\phi)} {3H_0^2}, \qquad \Om_{M0}^{}=\d {8\pi G\rho_0}{3H_0^2}.
\ee
In terms of new variable, Eq.(\ref{w}) becomes
\be 
w= \d {\chi ^2 -2\chi b_0 -2v(\varphi)}{\chi ^2 -2\chi  b_0  +2  
 v(\varphi) }. 
\ee
The initial values of integration are assigned to fit the present observation 
data \cite{WMAP,ex}: 
\be
h_0=1,\qquad &\Om_{{\rm DE}0}^{}:=\left .\left(\frac 1 2 \chi^2 -b_0\chi +v \right)\right |_0 =0.73,&\qquad  \Om_{M0}^{} = 0.27,\\
w_0=-1.33,  \qquad &\chi_0^{}(\chi_0^{} -2b_0) =-0.24\omits{09},& \qquad v_0^{} =0.85\omits{045}.
\ee

Now, we consider several examples.  The first example is the 
universe with a quadratic potential 
\be \label{potential}
V(\phi) = \d 1 2 m^2 \phi^2 .
\ee 
We may further take $b_0 =-0.6$ and 
\be
w_1=\left .\d {dw}{dz}\right |_0 =1,
\ee
then $\chi_0 \approx =-0.255$, $d\varphi/dz|_{0}^{}=-\chi_0\omits{=0.254891}$,
\be
v'(\varphi)|_{z=0}^{}=\d {\frac {1} 4 [\chi_0(\chi_0-2b_0)+2v_0]^2-6v_0(\chi_0-b_0)^2} {
\chi_0^2(\chi_0-2b_0) +2v_0(\chi_0-b_0)}=-0.117\omits{652},
\ee
$\varphi_0=-14.6$ and $\frac {m^2}{H_0^2}=8.0\times 10^{-3}$.  
FIG. 1 gives the evolution of $w$ for a flat universe with a quadratic potential.  It is obvious that the equation of state
across $w=-1$ at $z\approx 0.2$, which is consistent
with the SNe Ia observation \cite{HC} and that the transition redshift 
between the accelerating and decelerating phases occurs at $z \approx 0.5$, which is lower than the observational constraint
on the transition redshift interval $0.6<z<1.7$ 
\cite{Perl99,Riess98,Riess01}.  The limit of $w$ at large $z$ is 1. 
\begin{figure}[htpb]
\includegraphics[scale=0.8]{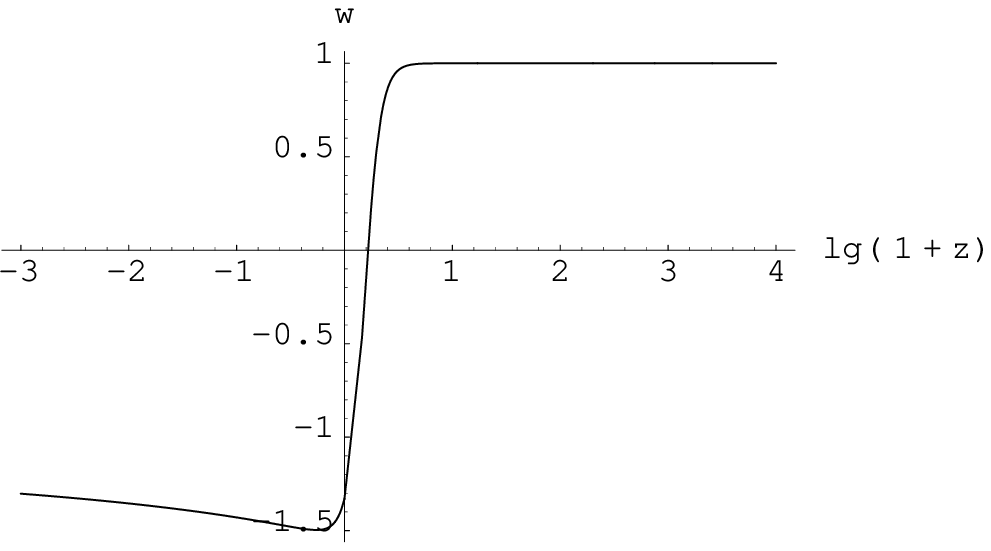}
\includegraphics[scale=0.8]{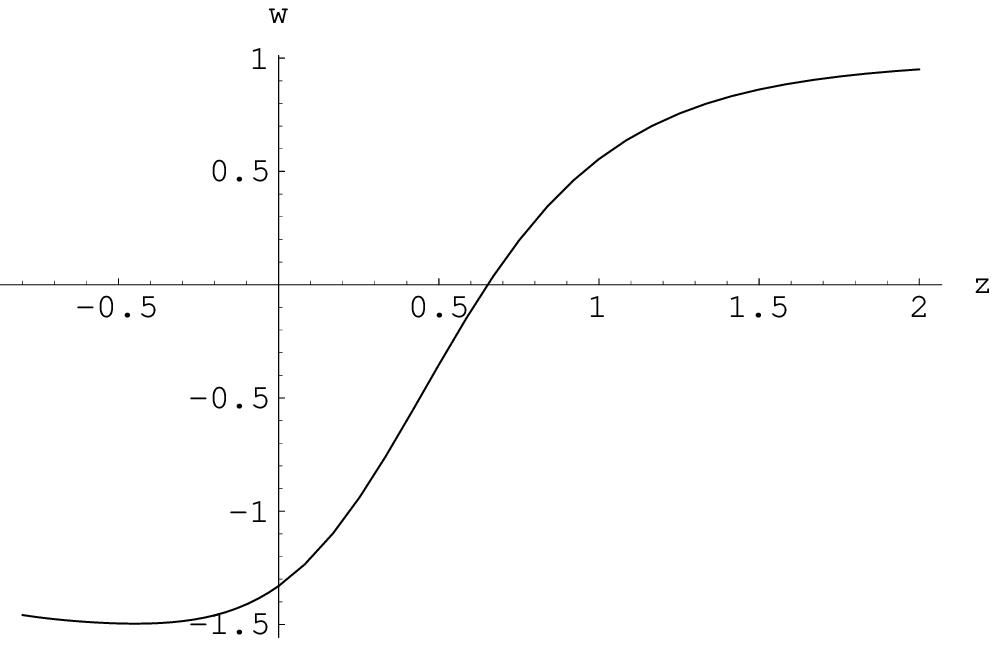} 
\caption{The evolution of $w$ for a flat universe with a quadratic potential with the initial values $w_0=-1.33$, $w_1=1$, $b_0=-0.6$,  $\varphi_0=-14.6$, $\frac {m^2}{H_0^2}=8.0\times 10^{-3}$, and $\Om_{M0}=0.27$. The right figure
is the expanded view of the left figure between $z=-0.8$ and $z=2$.}
\end{figure}
For $b_0=0.6$ the behavior of $w$ is similar to the case $b_0=-0.6$. 
It can be shown that $w$ has the similar behavior for the quartic potential 
\be 
V(\phi) = \d 1 4 \la \phi^4 ,
\ee
and the inverse power-law potential 
\be 
V(\phi) = V_0 \phi^{-\al} .
\ee

The next example is the universe with the potential 
inspired from supergravity \cite{BM} ,
\be \label{supgravpot}
V(\phi) = V_0 \phi^{-\al}e^{4\pi G \phi^2},
\ee
where $\al \geq 11$ is a constant.  For the potential, we may take
$\al=12$, $b_0=-0.62$, $w_1=1.47$, thus $v_0'(\varphi)=-0.129\omits{061}$, 
$\varphi_0=-2.0\omits{2545}$, $(\frac {3}{8\pi G})^{5} \frac {V_0}{H_0^2} =8.6\omits{179}$.
Fig. 3 gives the evolution of $w$ for the model.
\omits{
\be
v(\varphi)|_{z=0} &=& \left (\d {8\pi G} 3 \right )^{1+\al/2}\d {V_0}{H_0^2}\varphi^{-\al}|_{z=0}  =0.85045 \\
v'(\varphi)|_{z=0} &=& -\left (\d {8\pi G} 3 \right )^{1+\al/2}\d {\al V_0}{H_0^2}\varphi^{-\al -1}|_{z=0}=0.273784
\ee
$\phi_0=-0.85045\al/0.273784=-3.10628\al$ and $\left (\d {8\pi G} 3 \right )^{1+\al/2}\d {V_0}{H_0^2}=0.85045*(-3.10628\al)^\al$. } 
The equation of state across $w=-1$ at $z\approx 0.2$ and the 
transition redshift between the accelerating and decelerating phases 
occurs at $z \approx 1.6$, which is consistent with the SNe Ia observation \cite{Perl99,Riess98,HC,Riess01}.  However, the dependence of $w$ on $z$ is not 
monotonic.  $w$ becomes less than $-1$ between $z\approx 0.7$ and $z\approx 1.3$ again.  The universe experienced another accelerated expansion around $z\approx 3.8$.
\begin{figure}[hptb]
\includegraphics[scale=0.8]{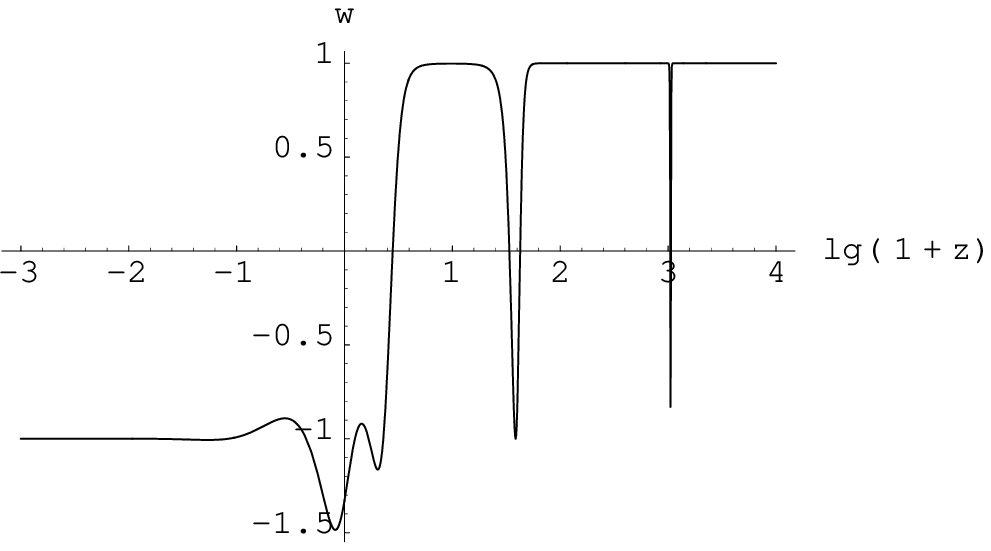}
\includegraphics[scale=0.8]{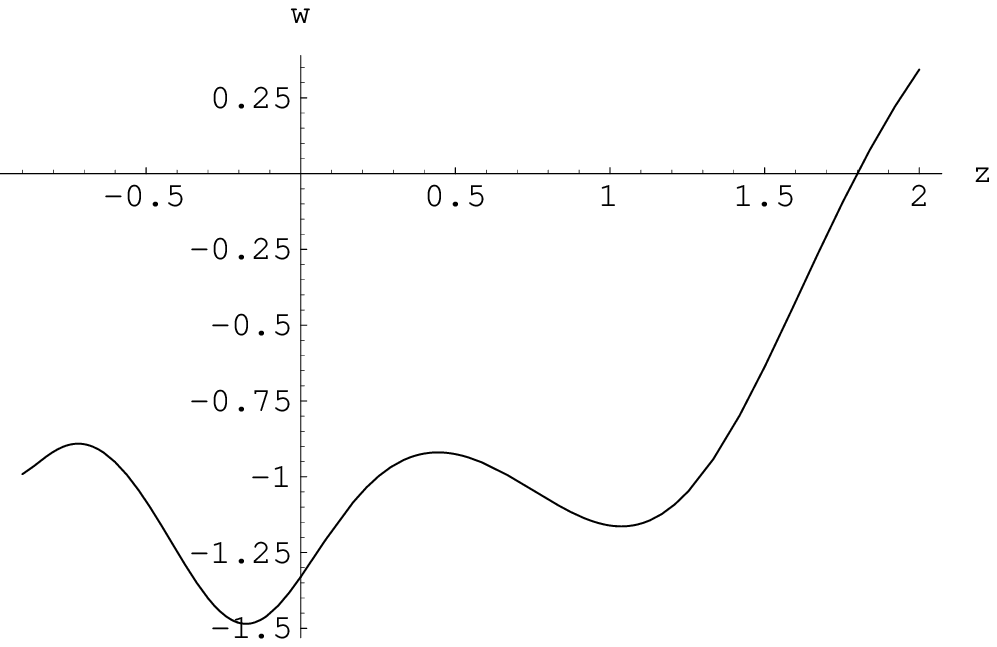}
\caption{The evolution of $w$ for a flat universe with a quadratic potential with the initial values $w_0=-1.33$, $w_1=1.47$, $b_0=-0.62$, $\al=12$ and $\Om_{M0}=0.27$. The right figure
is the expanded view of the left figure between $z=-0.9$ and $z=2$.}
\end{figure} 

\section{Concluding remarks}

The single-dynamical-scalar-field model of dark energy 
may explain the parameter $w$ 
in the effective equation of state crossing $-1$ during the 
evolution of the universe, as long as a fixed background vector field is
introduced.  For some potentials, the parameter $w$ crossing $-1$ and 
transition from decelerating to accelerating occur at $z\approx 0.2$ and $z\approx 1.7$, respectively, which is consistent with the observations. 
The dependence of the parameter $w$ on the redshift $z$ is far from
the linear relation.  Thus, it is needed to re-analyze the observation
data based on more general form of $w(z)$.  By the way, without the 
introduction of oscillation potential, the universe may also exhibit 
some oscillation behaviors. 

The present model is in the almost standard framework of physics, {\it e.g.} 
in general relativity in 4-dimension.  There does not exist phantom in the
model, which 
will lead to theoretical problems in field theory.  Instead, 
an {\it a priori} vector field, $A_\mu$, is introduced, which is selected in an 
unspecified way by the cosmological background.  It will result in
the violation of the Lorentz invariance and might be contributed
by the collective effects of the vacuum polarization of some fundamental 
fields.  It is needed to further
explore the physical meaning of such a vector field.

\section*{Acknowledgements}
The work is supported in part by the National Natural Science Foundation 
of China under the grant No. 90403023 and 10375087.

\end{document}